\documentstyle[aps,prb,epsfig,multicol]{revtex}
\begin{document}
\title{Computer simulations of the mechanism of thickness selection in polymer crystals}
\author{Jonathan P.~K.~Doye}
\address{University Chemical Laboratory, Lensfield Road, Cambridge CB2 1EW, UK}
\date{\today}
\maketitle
\begin{abstract}
In this paper I describe the computer simulations that I have performed to critically
examine the Lauritzen-Hoffman and the Sadler-Gilmer theories of polymer crystallization.
In particular, I have computed the free energy profile for nucleation of a new crystalline layer
on the growth face to compare with that assumed by the Lauritzen-Hoffman theory, 
I have analysed the mechanism of thickness selection in a multi-pathway model in which
some of the constraints in the Lauritzen-Hoffman theory are relaxed, and I have re-examined
the model used by Sadler-Gilmer. These investigations have lead to a mechanism of thickness
selection of lamellar polymer crystals that differs from the two theories that I set out to examine.

\end{abstract}
\begin{multicols}{2}

\section{Introduction}

In 1957 Andrew Keller reported that polyethylene formed chain-folded lamellar crystals from solution.\cite{Keller57a} 
This discovery was followed by the confirmation of the generality of this morphology---lamellar 
crystals are formed on crystallization from both solution and the melt\cite{Toda93}  for 
a wide variety of polymers---and the basic phenomenological laws describing such properties 
as the thickness and growth rate.\cite{Keller68a,Keller96a} 
In particular, the crystal thickness, $l$, has been found to be inversely
proportional to the supercooling,\cite{Keller58,Barham85} which is interpreted as 
resulting from $l$ being slightly larger than $l_{\rm min}$, the minimum thickness for which
a lamellar crystal is stable with respect to the solution or melt, i.e.\ 
$l=l_{\rm min}+ \delta l$, where $\delta l$ is small. 

Surprisingly, however, no theoretical consensus has yet been reached as to the mechanism of
this seemingly simple behaviour. In particular, two of the most well-known theories---the 
Lauritzen-Hoffman (LH) surface nucleation theory\cite{Lauritzen60,Hoffman76a,Hoffman97} 
and the Sadler-Gilmer (SG) entropic barrier model\cite{Sadler84a,Sadler86a,Sadler87d,Sadler88a}---present
very different explanations of thickness selection.\cite{Armistead92a} 
Of course, in such a situation, one would like to determine which of the theories, if any, is closest to the truth.
There are two aspects to such a task. Firstly, the predictions of the theories should be 
critically compared with experimental results. In the case of polymer crystallization 
both the LH and SG theories are able to reproduce the basic behaviour: 
the observed temperature dependence of the thickness and the growth rate. 
Additionally, Hoffman and coworkers have 
further developed 
the surface nucleation approach 
in order to explain some of the more detailed behaviour of crystallizing polymers, for example
the regime transitions in the growth rate.\cite{Hoffman97} 
However, this comparison does not conclusively favour one of the theories. 
This situation illustrates the fact that although consistency with experiment is 
an important first hurdle for any theory, it 
does not automatically imply the correctness of a theory. 
There may be a number of different ways of generating a particular experimental law. Furthermore, 
the number of parameters in a complex theory may give the theory sufficient plasticity 
to fit a wide variety of scenarios. 

Secondly, it is important that the assumptions of a theory, particularly those
about the microscopic mechanisms, are critically examined. However, in the case of polymer 
crystallization this task is very difficult to achieve experimentally. By addressing this gap,
computer simulations can potentially play an important role in this field.
Such simulations could range from examining simple models
to performing realistic atomistic simulations of the crystal growth process.
The former could allow the effects of relaxing some of the theoretical assumptions to be determined
and the latter could provide a detailed molecular picture of the growth process. 
Indeed, there has been an increasing number of computational studies pursuing these 
aims.\cite{Yamamoto97,Yamamoto98,Chen98,Liu98a,Toma98}
In this paper I will review my efforts in this direction\cite{Doye98f,Doye98d,Doye99b,Doye99d,Doye99e} 
and hope to illustrate the positive role that computer simulations can play in 
helping to understand polymer crystallization.
In particular, the aim of my simulations has been to critically examine the LH and SG theories.

\section{Free Energy profiles}

\begin{center}
\begin{figure}
\epsfig{figure=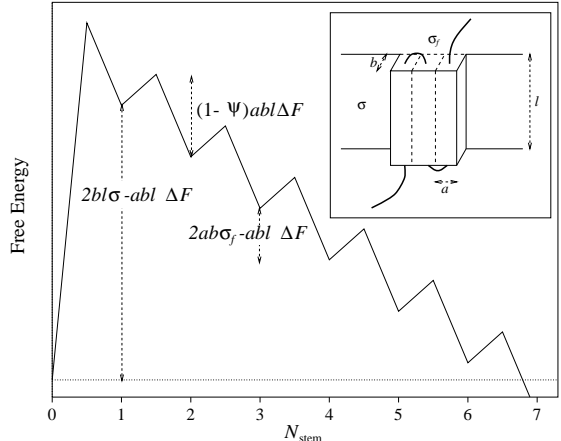,width=8.2cm}
\begin{minipage}{8.5cm}
\caption{\label{fig:SNfree} The free energy profile for the nucleation and growth of a 
new layer assumed by the LH theory. The inset is a schematic representation of a 
configuration with three stems deposited.
}
\end{minipage}
\end{figure}
\end{center}

In the LH theory the growth of a new layer is modelled as the deposition of a succession
of stems (straight sections of the chain that traverse the growth face) along the growth
face from an initial nucleus, where the length of each stem is the
same as the thickness of the lamella. The inset of Figure \ref{fig:SNfree} illustrates the geometry
of this mechanism. To analyse the kinetics of growth, 
a thermodynamic description of the nucleation and growth of a new layer is first required. 
The free energy of a configuration with $N_{\rm stem}$ complete stems is taken to be
\begin{equation}
A(N_{\rm stem})= 2 bl \sigma + 2 (N_{\rm stem}-1) a b \sigma_f - N_{\rm stem} a b l \Delta F,
\end{equation} 
where $a$ and $b$ are the width and depth of a stem, $l$ is the thickness of the lamella, 
$\sigma$ is the lateral surface free energy, $\sigma_f$ is the fold surface free energy, and $\Delta F$
is the free energy of crystallization.
The first term corresponds to the free energy of the two lateral surfaces created on 
the deposition of the first stem and is proportional to $l$. 
The second term is the free energy
of the new fold surface created on the deposition of subsequent stems. 
It is then assumed that at the barrier between configurations with different numbers of stems
all the new surfaces have been created and that a fraction $\Psi$ of the 
free energy of crystallization is released. This then gives the LH free energy profile
that is illustrated in Figure \ref{fig:SNfree}.

From this free energy profile, $S(l)$, the flux over the barrier, can be obtained. The observed 
crystal thickness is then taken to correspond to the average
\begin{equation}
\overline{l}=\int_{l_{\rm min}}^\infty l S(l) dl.
\label{eq:lave}
\end{equation}
This average thickness is close to the value of $l$ at the maximum in $S(l)$, which
in turn is close to, but slightly above $l_{\rm min}$, thus 
reproducing the observed behaviour of $l$.
The maximum in $S(l)$ is the result of two competing factors. 
The free energy barrier for deposition of the first stem increases with $l$, 
thus making the growth of thick crystals prohibitively slow. 
However, as $l_{\rm min}$ is approached from above, the thermodynamic driving 
force for crystallization goes to zero.

It is important to note that by integrating over $l$, Equation (\ref{eq:lave})
assumes that there are crystals with all values of $l$ greater than $l_{\rm min}$ which all grow
with constant thickness and contribute to the average $\overline{l}$. Those crystals with
a thickness close to the maximum in $S(l)$ dominate this ensemble and contribute more 
to Equation (\ref{eq:lave}) because of their rapid growth.
As was realized by Frank and Tosi,\cite{Frank60} the results of experiments 
where the temperature is changed during crystallization argue against such an ensemble.
The temperature jumps give rise to steps on the lamellae, showing that a crystal need
not necessarily grow at constant thickness.\cite{Bassett62,Dosiere86a} 

We will come back to this issue later, but in this section we focus on the LH 
free energy profile. In particular, we compare this theoretical profile 
with ones computed from simulations of a simple polymer.\cite{Doye98f}
In our model the polymer is represented by a self-avoiding walk on a simple cubic lattice.
There is an attractive energy, -$\epsilon$, between non-bonded polymer units on
adjacent lattice sites and between polymer units and the surface, 
and an energetic penalty, $\epsilon_g$, for kinks (or `gauche bonds') 
in the chain. The parameter $\epsilon_g$ determines the stiffness of the chains.
In our simulations we have included a surface which represents the growth face of a polymer crystal.

\begin{center}
\begin{figure}
\epsfig{figure=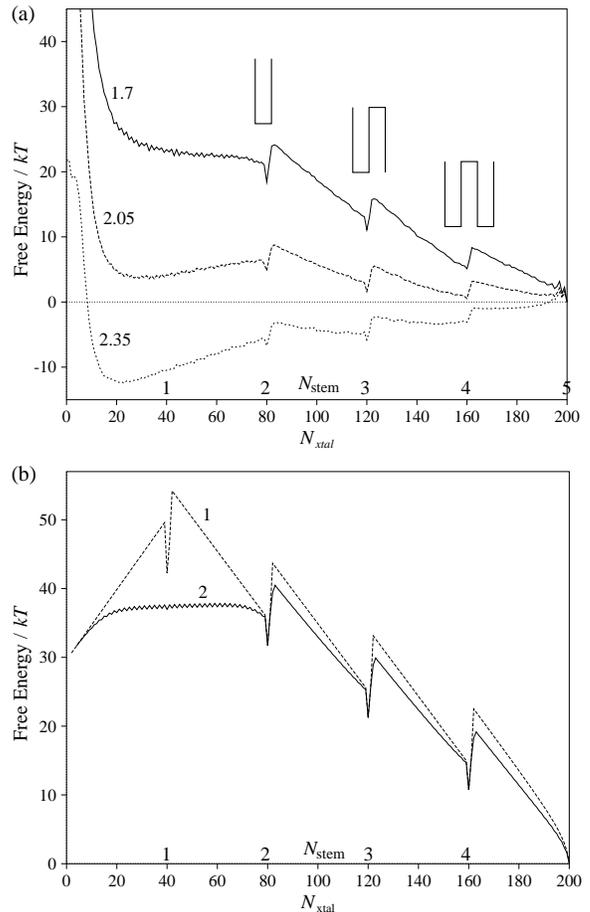,width=8.2cm}
\begin{minipage}{8.5cm}
\caption{\label{fig:MCfree}
Free energy profiles for the formation of a target crystal with 5 stems of length 40 units
and adjacent reentry of the folds.
In (a) the profiles have been calculated from simulation,
the labels give the temperature, and example configurations along the pathway have
been illustrated..
In (b) the profile has been calculated at $T=2.75\,\epsilon k^{-1}$ using Equation (\ref{eq:simple})
for pathways which allow one (1) or two (2) incomplete stems, as labelled. $\epsilon_g=4\epsilon$.
}
\end{minipage}
\end{figure}
\end{center}

To follow the crystallization of the polymer on the surface, we need to define an 
order parameter which determines the degree of crystallinity. We use $N_{\rm xtal}$, the
largest fragment of the polymer with the structure of a target crystalline configuration.
In our case, we examine the crystallization of a 200-unit chain into a structure with 5 stems 
of length 40 units. 
In order to compare with the theoretical profiles we have to constrain the 
other $N-N_{\rm xtal}$ units in the chain to be disordered. The simulations were carried out
using configurational-bias Monte Carlo,\cite{Siepmann92a} and the umbrella sampling 
technique\cite{Torrie77} was used to calculate the free energy profiles.

The free energy profiles that we obtained are shown in Figure \ref{fig:MCfree}a. They show 
the expected temperature dependence: at low temperature the crystal is most stable and at high 
temperature the disordered state is most stable. 
Note that the value of $N_{\rm xtal}$ for the disordered state is non-zero, because
the disordered polymer is adsorbed on the surface.
The adsorbed polymer is bound to have some short straight
sections that qualify as crystalline by the definition of $N_{\rm xtal}$.
The free energy profiles also have a sawtooth structure resembling that of the theoretical profile.
The barriers occur immediately after the previous stem has been completed, and
correspond to the formation of a new fold. They are followed by a monotonic 
decrease in energy as this new stem grows to completion.
In the language of the LH theory $\Psi({N_{\rm stem}\rightarrow N_{\rm stem}+1})\approx 0$ for $N_{\rm stem}\ge 2$.
However, there is no feature in the simulation profiles that corresponds to the formation of the first fold. 
This is because the initial nucleus is not a single stem, 
but two stems connected by a fold that grow simultaneously. Such a possibility had
previously been suggested by Point.\cite{Point79b}

Confirmation of a two-stem nucleus comes from a simple model calculation of the free energy profile.
We can write the free energy as 
\begin{equation}
A(N_{\rm xtal})=A_{\rm coil}(N-N_{\rm xtal}) + 
                kT \sum \exp \left( -E_{\rm xtal}/kT\right),
\label{eq:simple}
\end{equation}
where the sum is over all possible crystalline configurations which are $N_{\rm xtal}$ units long,
$E_{\rm xtal}$ is the energy of the crystalline configuration, and
$A_{\rm coil}$ is the free energy of an ideal two-dimensional coil. 
The resulting profile is very similar to the simulation profile (Figure \ref{fig:MCfree}b). 
In particular, there is no feature due to the formation 
of the first fold. However, when we force the initial nucleus 
to be a single stem by restricting the sum in the above equation to only 
those crystalline configurations with one incomplete stem, 
a free energy barrier associated with the formation of the first fold appears.
The reason for the preference for a two-stem nucleus is simply energetic. 
For $N_{\rm xtal}> 4\epsilon_g/\epsilon +2$ the two-stem nucleus is lower in energy 
because of the interaction between the two stems. Our simulations
were performed on a surface that was infinite. Whether a two-stem nucleus would be
expected, when, as with a lamellar crystal, the thickness of the growth face 
is finite, depends upon how this critical size compares to the thickness of the lamella.

It can be clearly be seen from Figure \ref{fig:MCfree}b that the two-stem nucleus
significantly reduces the nucleation barrier. In particular, it will no longer be
proportional to $l$. This has significant implications for the LH theory given
the key role played by this initial free energy barrier in
constraining $\overline{l}$ to a value close to $l_{\rm min}$.

Before we move on we should make a number of comments.
First, the polymer model is very simple, and although there is no obvious reason why
the thermodynamic reasons behind the two-stem nucleus should not also apply to
a real polymer, there may be factors that are not included in our model that come into play.

Second, the profiles reflect our choice of order parameter. As we monitor crystallization
unit by unit, during the growth of the first two stems the lateral surface energy is paid for
at the same time as the free energy of crystallization is released. Therefore, in this
size range $\Psi$ is effectively equal to 1, albeit with the possibility of a two-stem nucleus.
Hoffman, however, advocates a $\Psi$=0 version of the LH theory---it has the advantage
that it avoids a `$\delta l$-catastrophe' (a divergence of the lamellar thickness) at 
large supercooling---in which he postulates that prior to crystallization 
an aligned physisorbed state is formed that has lost its entropy but not yet gained
the free energy of crystallization.\cite{Hoffman97} Such a state cannot occur in our lattice model because 
there is no difference in the interaction with the surface for a disordered
chain adsorbed on the surface and a crystalline layer. In an off-lattice model the interaction
energy for the crystalline layer would be greater because the stems would fit into the
grooves provided by the stems of the previous layer. 

\begin{center}
\begin{figure}
\epsfig{figure=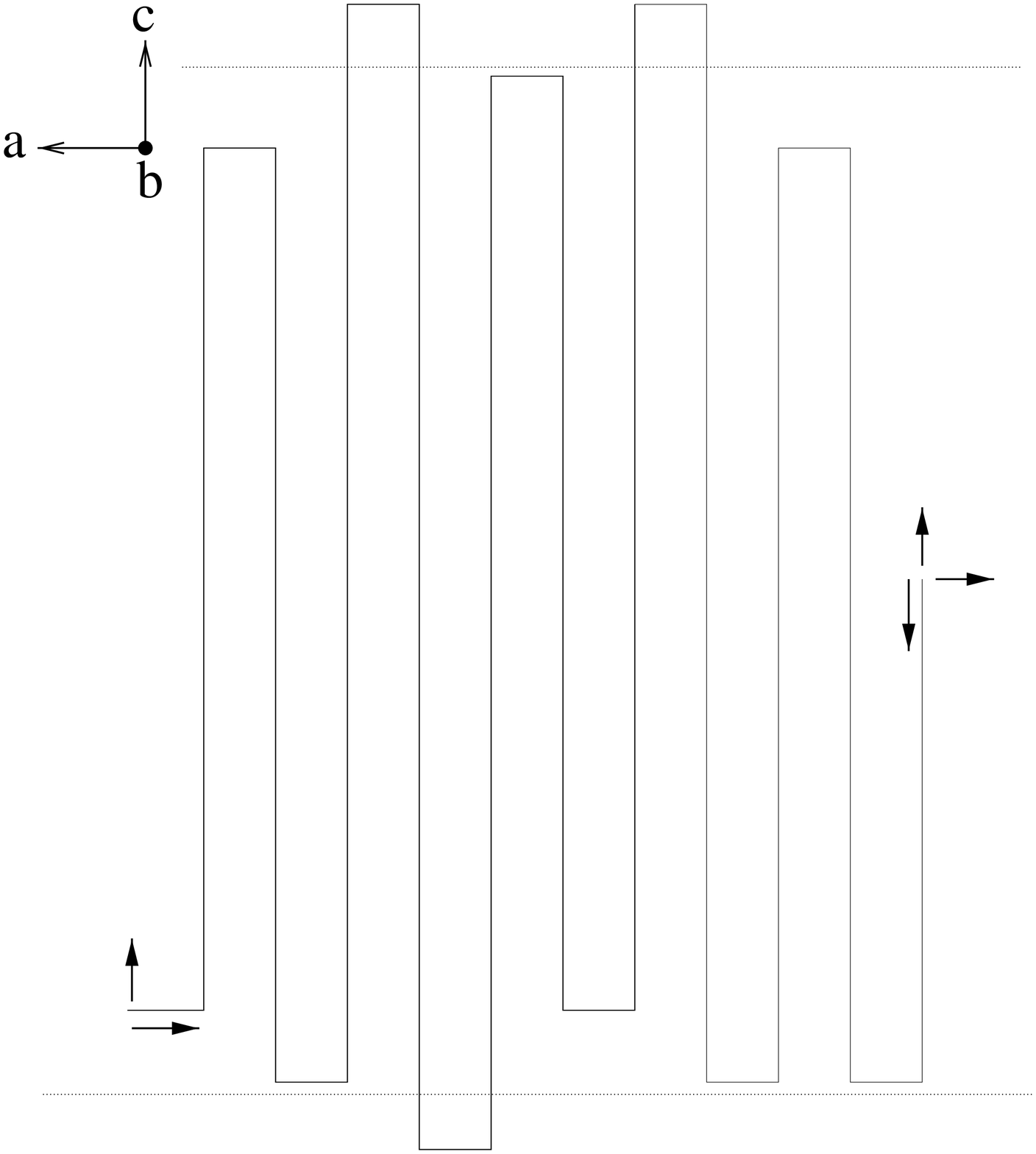,width=5.2cm}
\begin{minipage}{8.5cm}
\caption{\label{fig:MPmoves}
An example crystalline configuration during the growth of a new layer on the surface of 
the growth face in the multi-pathway model. The arrows indicate the five possible moves 
at the next step in the kinetic Monte Carlo simulation. The dotted lines show the edges of the
growth face.
}
\end{minipage}
\end{figure}
\end{center}

Third, a good order parameter must pass continuously
through intermediate values when the system goes between two states. 
However, one can imagine a number of mechanisms by which this criterion for $N_{\rm xtal}$ is broken. 
For example, in a realistic simulation of the surface crystallization of a long alkane into a once-folded 
configuration, the chain first formed non-adjacent crystalline stems connected by 
a loose fold which then came together by the propagation of a defect through one of the stems.\cite{Liu98a}
Another possibility that has been observed in simulations is the formation of crystallites in 
different portions of a chain that subsequently coalesce to form a single crystallite.\cite{Yamamoto97,Toma98}

\section{A multi-pathway model}

In the previous section, in order to compare the LH free energy profile with those from 
simulation, we had to constrain the $N-N_{\rm xtal}$ units
not having the target structure to be disordered. If we had not done this, at temperatures
where the crystal is most stable the rest of the chain would have formed a crystalline configuration
with stem lengths 
different from the target configuration. This naturally raises questions about 
the LH assumption that the stems in a new layer must all have the same thickness as the previous layer.
In this section, we examine the effects of relaxing some of the LH assumptions by 
studying a model in which the stems grow unit by unit and the length of a stem is 
unconstrained.\cite{Doye98d,Doye99b}
We term it a multi-pathway model because it can take into account the many possible ways that a new 
crystalline layer can form.

This idea is not new. Frank and Tosi,\cite{Frank60} Price\cite{Price61} 
and Lauritzen and Passaglia\cite{Lauritzen67b} considered models where the stem length is not always constant, and 
Point,\cite{Point79a} and DiMarzio and Guttman\cite{DiMarzio82a} studied models where the stems could grow 
unit by unit. All these studies were performed at a time when computational resources were much less,
so approximations and simplifications had to be made in order to render the models tractable.
The natural way to solve such problems, though, is through the use of computational 
techniques, such as kinetic Monte Carlo.
However, the only applications of computational methods to this problem were in a short note by
Point\cite{Point79c} and the continuation of this work in the PhD thesis of Dupire.\cite{Dupirelong}
Some of the results presented in these earlier studies are similar to those we report here.

In our model we grow a single new crystalline layer by the successive growth of stems across
a surface that represents the growth face of a polymer crystal. The polymer interactions are
the same as used in the previous section, and we only model the crystalline portion of the polymer 
explicitly---the rest is assumed to behave like an ideal coil. An example configuration is illustrated 
in Figure \ref{fig:MPmoves} along with possible changes of configuration. These changes can only 
occur at the ends of the crystalline portion, and are selected using the kinetic Monte Carlo 
algorithm, in which a move is chosen with a probability proportional to the rate for that process.

\begin{center}
\begin{figure}
\epsfig{figure=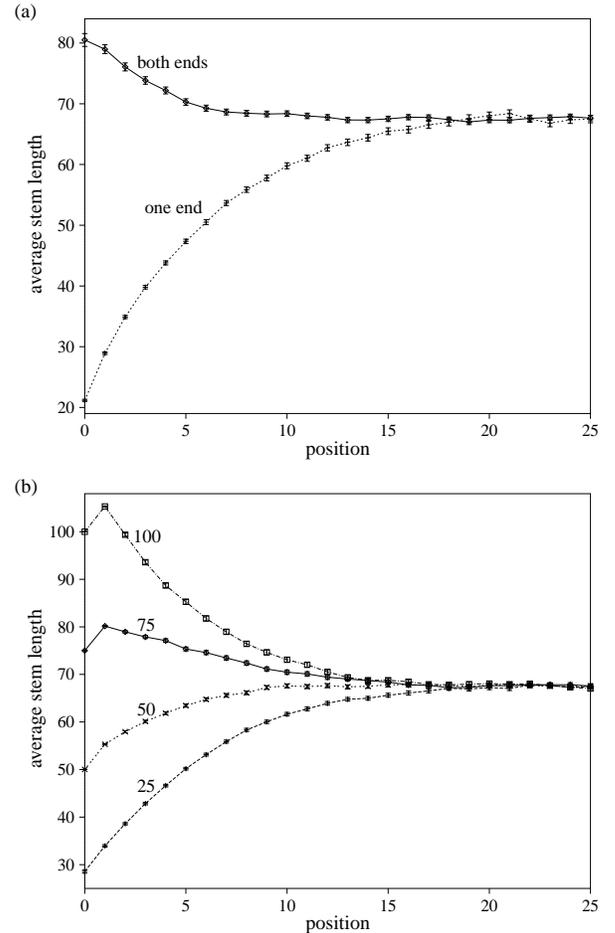,width=8.2cm}
\begin{minipage}{8.5cm}
\caption{\label{fig:MPnucleus}
The dependence of the average stem length on the distance of the stem from
(a) the initial nucleation site, and
(b) the centre of an initial crystal seed
for growth on an infinite surface at $T=2.75\,\epsilon k^{-1}$.
In (a) growth starts with a single polymer unit on the surface and we consider the cases 
where growth is allowed at one end or both ends of the crystalline configuration.
In (b) the crystal seeds are 3 stems wide; the lengths of the stems in the seeds are
as labelled. $\epsilon_g=8\epsilon$.
}
\end{minipage}
\end{figure}
\end{center}

First, we shall examine the effect of the initial nucleus on the thickness of the layers grown. 
If the stem lengths are unconstrained and the initial nucleus is a single stem, one might
imagine that one way of reducing the large initial free energy barrier in Figure \ref{fig:SNfree}
(and achieving faster initial growth) would be for the stem length to increase gradually 
to its average value as crystallization progresses.
For this pathway, the lateral surface free energy is paid for `in installments' rather than
all initially. This is exactly what we observe when we force the initial nucleus to be a single
stem by only allowing growth from one end of the crystalline portion of the chain 
(Figure \ref{fig:MPnucleus}a). When a double-stem nucleus is allowed the initial growth is 
very different because there is now no longer a large initial free energy barrier to circumvent.
The most important thing to note from these results is that, contrary to the LH theory, 
the thickness of the inital nucleus does not determine the thickness of the layer. Further confirmation
of this can be obtained when we examine the growth from initial seed crystals.
Whatever the thickness of the initial seed the thickness of the growing crystal
converges to the same value (Figure \ref{fig:MPnucleus}b).
This implies that the thickness of a crystalline layer must be determined
by factors which are operating on the deposition of each stem and not those 
specific to the initial stems.

\begin{center}
\begin{figure}
\epsfig{figure=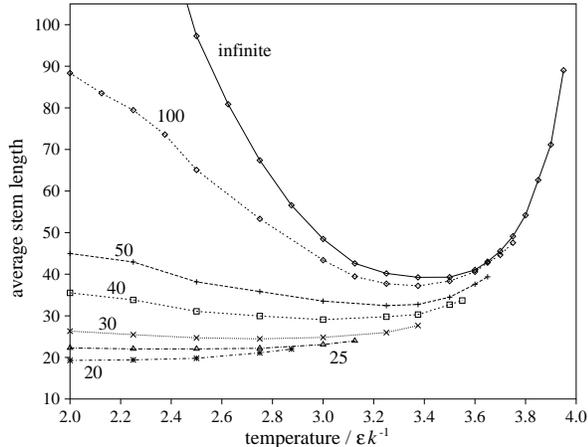,width=8.2cm}
\begin{minipage}{8.5cm}
\caption{\label{fig:MPthick}
The temperature dependence of the average stem length in a new crystalline layer for 
growth of a single layer on growth faces of different thickness, as labelled.
}
\end{minipage}
\end{figure}
\end{center}

To determine what these factors might be, in Figure \ref{fig:MPthick}
we show how the thickness of a new layer depends on temperature. 
First, it is immediately obvious that the thickness
of a new layer is not necessarily the same as that of the growth face.
Second, all the curves increase as the temperature approaches $T_m$, the melting or dissolution
temperature, because of the rise of $l_{\rm min}$.
Third, the thickness also increases at low temperature, in this instance because it becomes 
increasingly difficult to scale the free energy barrier for forming a fold
and so on average the stems continue to grow for longer.
However, this rise is checked by the thickness of the growth face. 
It is unfavourable for the polymer to overhang the edge of 
the growth face because these units do not interact with the surface.

\begin{center}
\begin{figure}
\epsfig{figure=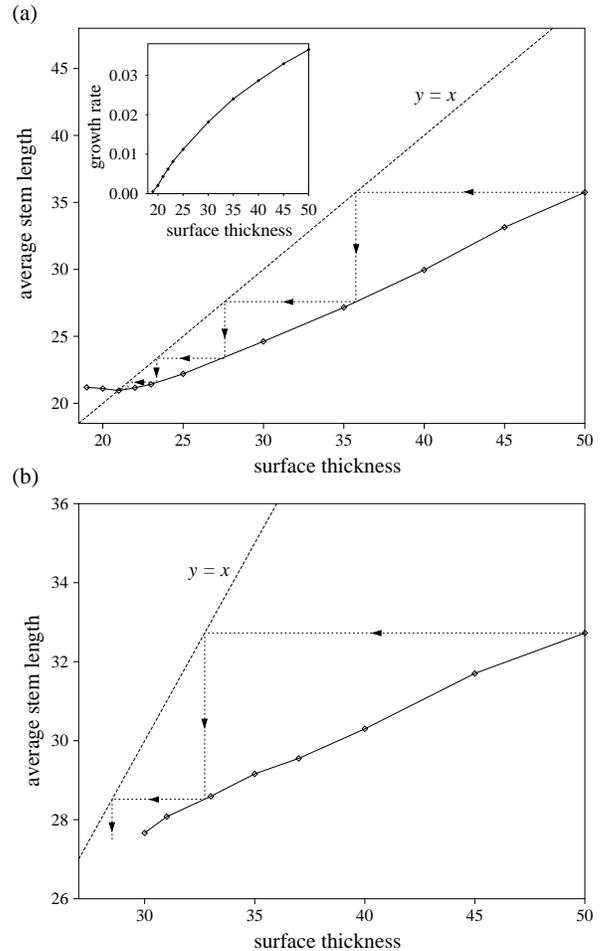,width=8.2cm}
\begin{minipage}{8.5cm}
\caption{\label{fig:MPattract}
The dependence of the average stem length in a new crystalline layer on 
the thickness of the growth face for the growth of a single layer
at (a) $T=2.75\,\epsilon k^{-1}$ and (b) $T=3.375\,\epsilon k^{-1}$.
The dotted lines show how the thickness changes on addition of successive
layers to a 50-unit thick surface.
The inset in (a) shows the growth rate for the new layer as a function of 
the thickness of the growth face.
}
\end{minipage}
\end{figure}
\end{center}

Figure \ref{fig:MPthick} only describes the growth of a single layer. However, 
as the thickness of the new layer is not generally the same as the thickness of the growth face, 
one needs to consider the addition of a succession of layers. If we assume that all the 
variations in the stem length within a layer are annealed out before a new layer begins to grow,
this can be achieved using Figure \ref{fig:MPattract}a, in which we have plotted for 
a single temperature the thickness of the new layer against the thickness of the growth face.
By following the dotted lines one can see what would happen for growth on a
growth face that is 50 units thick: the first layer is 36 units thick, 
the second 28, the third 23, \dots Thus, the thickness converges to the value 
$l^{**}$ at which the curve crosses $y=x$, i.e.\ to the point where the thickness of the 
new layer is the same as the previous, and then the crystal continues to grow at this thickness.
The mapping represented in Figure \ref{fig:MPattract}a is a fixed-point attractor.

\begin{center}
\begin{figure}
\epsfig{figure=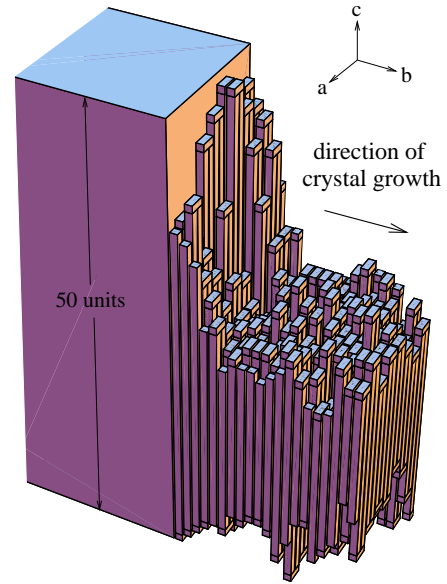,width=8.2cm}
\begin{minipage}{8.5cm}
\caption{\label{fig:MP3D}
Cut through a polymer crystal which was produced by the growth of
twenty successive layers on a growth face with a uniform thickness 
of 50 units at $T=2.0\,\epsilon k^{-1}$.
The stems are represented by vertical cuboids. The cut is 16 stems wide.
}
\end{minipage}
\end{figure}
\end{center}

A similar picture emerges if we explicitly perform simulations of multi-layer growth. 
Figure \ref{fig:MP3D} shows a cut through a typical configuration that results.
Within 5--10 layers the thickness of the crystal converges
to its steady-state value $l^{**}$ and then growth continues at that thickness.

The mechanism of thickness selection that occurs in our 
multi-pathway is at odds with the LH theory. 
It shows that it is inappropriate to compare the growth rates of crystals of different
thickness because the thickness has only one dynamically stable value for
which growth at constant thickness occurs.
The ensemble of crystals assumed by Equation (\ref{eq:lave}) is fictitious.
Furthermore, the growth rate of a new layer slows down
as $l^{**}$ is approached from above (inset of Figure \ref{fig:MPattract}a).
However, we should note that in some of the multiple-pathway studies\cite{Frank60,Lauritzen67b,Price61}
mentioned earlier, it was realized that stable growth can only occur at the one thickness where
a new layer has the same thickness as the previous. Since then this insight has for the most
part been neglected.

\begin{center}
\begin{figure}
\epsfig{figure=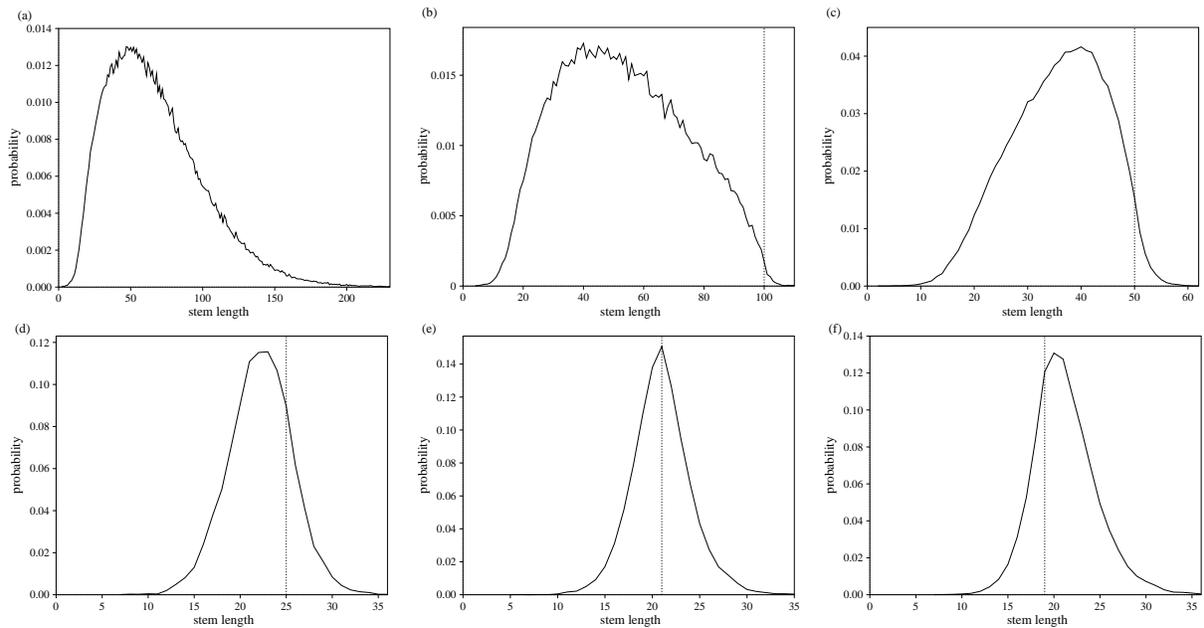,width=16.2cm}
\begin{minipage}{17.6cm}
\caption{\label{fig:MPpstem}
Probability distributions of the stem length for 
a new crystalline layer grown at $T=2.75\,\epsilon k^{-1}$
on a growth face of thickness: (a) $\infty$, (b) 100, (c) 50, (d) 25, (e) 21 and (f) 19.
The dashed vertical lines in the probability distributions are
at the thickness of the growth face.
}
\end{minipage}
\end{figure}
\end{center}

To analyse the reasons for the dynamical convergence of the thickness to the value $l^{**}$ 
we examine how the probability distributions for the stem length depend on the thickness of the
growth face (Figure \ref{fig:MPpstem}).
$l_{\rm min}$ places one constraint on the stem length; only a small
fraction of the stems can be shorter than $l_{\rm min}$
if the layer is to be thermodynamically stable.
The thickness of the growth face places the second constraint on the stem length;
it is energetically unfavourable for the polymer to extend beyond the edges of the growth face.
There is also a third weaker kinetic constraint on the stem length. 
At every step there is always a certain probability that a fold will be formed. 
Therefore, even in the absence of the second constraint, i.e.\ an infinitely
thick growth face, the probability distribution will decay exponentially 
to zero at large stem length (Figure \ref{fig:MPpstem}a).
Although, this effect prevents the thickness from ever diverging in 
a $\delta l$-catastrophe,\cite{Point79a,DiMarzio82a} it 
does not stop the thickness becoming very large.

When the growth face is significantly thicker than $l_{\rm min}$ there is a 
range of stem lengths between $l_{\rm min}$ and the thickness of the growth face
that are viable, and therefore the new layer will be thinner than the previous layer.
However, as the thickness of the growth face decreases, the probability distributions of the stem length
becomes increasingly narrow and the difference in probability between the stem length being greater
or less than the surface thickness diminishes. Finally, at $l^{**}$, as the thickness of the growth face 
approaches $l_{\rm min}$, the probability distribution become symmetrical about the surface thickness 
and the thickness of the new layer becomes equal to the thickness of the growth 
face (Figure \ref{fig:MPpstem}e).
When the thickness is less than $l^{**}$, the asymmetry of the probability distribution
is reversed (Figure \ref{fig:MPpstem}f).
It is, therefore, through the combined action of the two thermodynamic constraints
on the stem length that the thickness converges to a value close to $l_{\rm min}$.

\end{multicols}
\begin{multicols}{2}

The picture is not quite this simple at all temperatures. As the supercooling decreases,
it becomes increasingly unfavourable for a stem to overhang the edge of the growth face. 
Indeed, for sufficiently small supercoolings 
the probability distribution for the stem length never becomes symmetrical about the
thickness of the growth face, not even when the thickness of the growth face is 
close to $l_{\rm min}$. This situation is illustrated in Figure \ref{fig:MPattract}b.
After the growth of two layers on a 50-unit thick surface, 
the crystal stops growing because the outer layer is too thin for a new layer to form.
For these supercoolings, as in the SG model, the rounding of the crystal profile inhibits growth. 
To overcome this barrier requires a cooperative mechanism
whereby a new layer takes advantage of (and then locks in)
dynamic fluctuations in the outer layer to larger thickness.
However, unlike the SG model, the current model has no interlayer dynamics---we attempt
to grow a new layer on an outer layer that is static---and so growth stops.
Despite this it is clear that if this interlayer dynamics could be included, it would
again lead to steady-state growth close to $l_{\rm min}$.

\begin{center}
\begin{figure}
\epsfig{figure=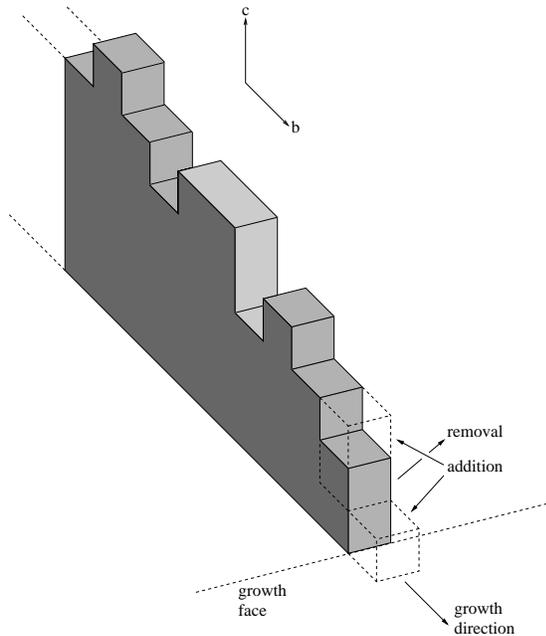,width=7.2cm}
\begin{minipage}{8.5cm}
\caption{\label{fig:SGmoves}
A schematic picture of a two-dimensional slice (perpendicular to the
growth face) through a lamellar polymer crystal which forms the basis of the two-dimensional
version of the Sadler-Gilmer model.
The three possible changes in configuration allowed by the model are shown (the dashed lines
represent the outline of the possible new configurations).
}
\end{minipage}
\end{figure}
\end{center}

We should note that this cessation of growth was also found in the model of Frank and
Tosi at low supercoolings.\cite{Frank60} Lauritzen and Passaglia were also aware of this effect, but they
introduced an {\it ad hoc} energetic term in their rate constants to prevent it.\cite{Lauritzen67b}
However, in the restricted equilibrium model of Price this effect was absent.\cite{Price61} In this study 
each new layer, but not the crystal as a whole, was allowed to reach equilibrium and so the kinetic
constraint on the stem length is absent.

Finally, we should note that our multi-pathway model is not parameter-free, 
and that, like most other models of polymer crystallization (including the LH 
theory\cite{Hoffman76a}, the SG model\cite{Doye99d,Goldbeck92a,Goldbeck94a} and
the earlier multi-pathway models\cite{Frank60,Price61,Lauritzen67b}), 
for some choices of parameters (not those used here) the lamellar thickness begins to increase at 
sufficiently large supercooling.\cite{Doye99b} This effect occurs because the large driving force for
crystallization at large supercoolings reduces the effect that the thickness of the
growth face has in constraining the stem lengths.

\section{The Sadler-Gilmer model}

In this section we re-examine the model used by Sadler and Gilmer in order 
to see whether the mechanism of thickness selection that we found in the 
previous section for our multi-pathway model also occurs in the SG model.
Sadler and Gilmer interpreted this model in terms of an entropic barrier. In 
particular, they argued that the rounding of the crystal profile gives rise
to an entropic barrier, which can only be surmounted by a fluctuation to a
squarer profile before growth can continue. As this barrier increases with lamellar
thickness it constrains the thickness to a value close to $l_{\rm min}$.
However, we shall not dwell on this interpretation here, but instead direct
the interested reader to a critique of this argument in Ref.\ \onlinecite{Doye99d}.

\begin{center}
\begin{figure}
\epsfig{figure=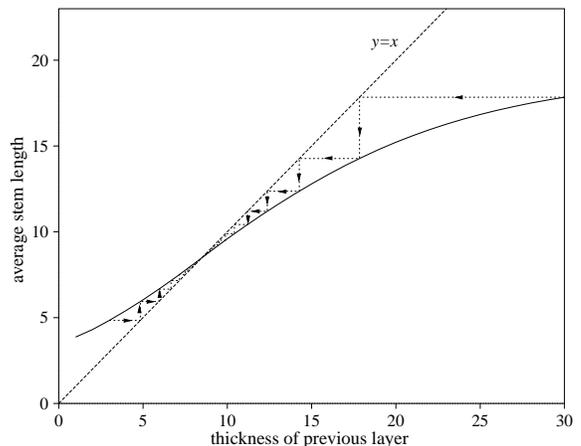,width=8.2cm}
\begin{minipage}{8.5cm}
\caption{\label{fig:SGattract}
The dependence of the thickness of a layer in the bulk of the crystal on the thickness of
the previous layer at $T=0.95T_m$.
The dotted arrowed lines show the thickness converging to the fixed point 
of the attractor from above and below.
}
\end{minipage}
\end{figure}
\end{center}

In the SG model the growth of a polymer crystal results from the attachment and detachment 
of polymer units at the growth face.  The rules that govern the sites at which these processes can occur
are designed to mimic the effects of the chain connectivity.
In the original three-dimensional version of the model,
under many conditions the growth face is rough and the correlations between stems in
the direction parallel to the growth face are weak.\cite{Sadler84a,Spinner95}
Therefore, an even simpler two-dimensional version of the model was developed in which
lateral correlations are neglected entirely, and only a slice through the polymer crystal perpendicular 
to the growth face is considered.\cite{Sadler86a,Sadler88a}

\begin{center}
\begin{figure}
\epsfig{figure=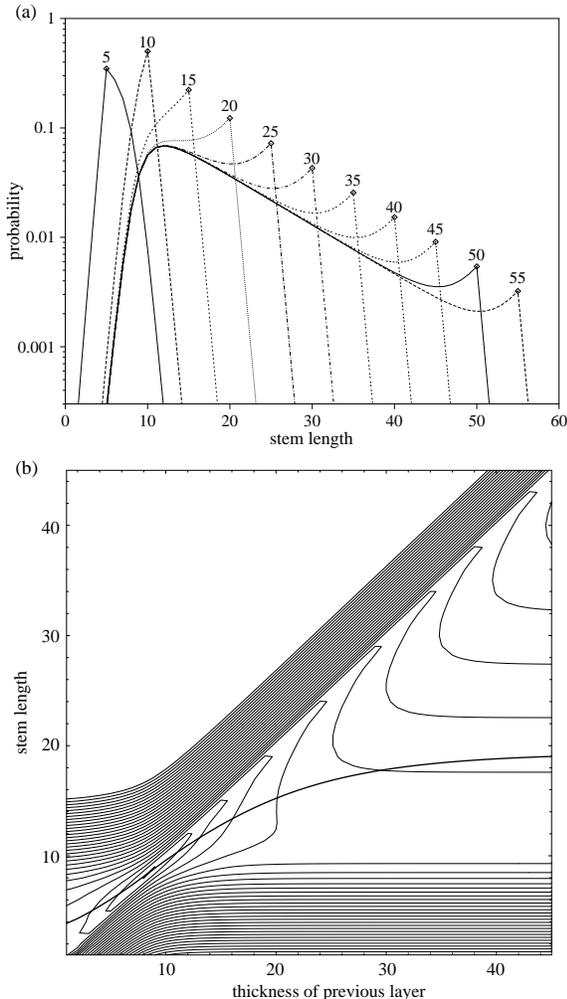,width=8.0cm}
\begin{minipage}{8.5cm}
\caption{\label{fig:SGpstem}
(a) Probability distributions of the stem length in the bulk of the crystal given that the
previous layer has the labelled thickness. (b) A contour plot of the log of a series of
such probability distributions. (Only thirty contours are displayed. The probability 
continues to fall rapidly in blank top left-hand corner.) 
The fixed-point attractor has been overlaid on this figure to illustrate its connection 
to the probability distributions. $T=0.95T_m$.
}
\end{minipage}
\end{figure}
\end{center}

The geometry of the model is shown in Figure \ref{fig:SGmoves}.
Changes in configuration can only occur at the outermost stem and stems behind
the growth face are `pinned' because of the chain connectivity.
At each step, there are three possible changes in configuration: 
the outermost stem can increase in length, a new stem can be initiated and
a polymer unit can be removed from the outermost stem.
The model can be formulated in terms of a set of rate equations 
that can be easily solved by numerical integration.\cite{Sadler86a}

When we examine the dependence of the thickness of a layer on the previous, 
we again find a fixed-point attractor describing the convergence of the thickness 
to its steady-state value (Figure \ref{fig:SGattract}). 
Moreover, when we examine the probability distributions for the stem length we 
find evidence for the same three constraints as for the multi-pathway model (Figure \ref{fig:SGpstem}b). 
The weaker nature of the kinetic constraint is particularly clear from the much more rapid exponential 
decay of the probability for stems that extend beyond the growth face. 
The role played by the two thermodynamic constraints in the mechanism of thickness
selection is particularly clear from Figure \ref{fig:SGpstem}b. As the thickness of the growth 
face decreases the viable range of stem lengths decreases until the the thickness of the
growth face meets $l_{\rm min}$ at the fixed point.

\section{Discussion}

In this paper we have outlined evidence from computer simulations for a mechanism of 
thickness selection in lamellar polymer crystal that differs from the theories of 
Lauritzen and Hoffman, and Sadler and Gilmer. Instead, the mechanism has much more in common with
the results of earlier multi-pathway models.\cite{Frank60,Price61,Lauritzen67b}
We find that a fixed-point attractor which describes the dynamical convergence
of the crystal thickness to a value just larger than the minimum
stable thickness, $l_{\rm min}$.
This convergence arises from the combined effect of two constraints on the
length of stems in a layer: it is unfavourable for a stem to be shorter than
$l_{\rm min}$ and for a stem to overhang the edge of the previous layer.
It is encouraging to note that we find the same mechanism of thickness selection
operating in two models which make very different assumptions about the microscopic 
growth processes. This provides evidence of the generality of this mechanism,
and so suggests that, although the models described here have a very
simplified description of the microscopic dynamics, the physical principles
behind the mechanism could be general enough to apply to real polymers.

This mechanism of thickness selection is also consistent with experiments
where the temperature is changed during crystallization.\cite{Bassett62,Dosiere86a} 
The steps that result indicate that the thickness of the lamellar crystals
dynamically converges to the steady-state thickness for the new temperature by 
a mechanism similar to that which we observe in our simulations.
Furthermore, if the step profiles could be characterized with sufficient 
resolution by atomic-force microscopy, it may be possible to extract 
the fixed-point attractor of a real polymers. However, for a temperature decrease 
the step profiles may also reflect the rounding of the crystal edge and 
for a temperature increase the roughness of the fold surface.\cite{Doye99e}
Furthermore, any annealing mechanisms that operate could change the shape of
the step profile from its as-formed state.

Although the multi-pathway approach is, in some ways, an extension of the LH theory,
the removal of many of the LH constraints leads to significantly different behaviour.
In particular, our work undermines the LH assumptions that the initial nucleus 
determines the thickness of a layer, and shows that the approach embodied in Equation (\ref{eq:lave}) 
(i.e.\ a comparison of the growth rates of the crystals 
in an ensemble of crystals of different thickness all of which grow at constant thickness)
is inappropriate because crystals of arbitrary thickness do not necessarily continue 
to grow at that thickness.
Although our results lead us to question the thickness selection mechanism in the LH 
theory, other aspects of the nucleation approach may not be affected by our critique.
For example, the regime transitions are a result of the different functional dependence
of the growth rate on the nucleation rate and the substrate completion rate in
the different regimes.\cite{Hoffman97}

Recently, there have been a number of alternative theoretical proposals
that have made recourse to metastable phases.
Keller and coworkers suggested that crystallization of polyethylene 
could initially occur into the mobile hexagonal phase. These crystals would
then thicken until a critical thickness was reached at which a phase transition
to the orthorhombic phase would occur.\cite{Keller93a,Keller98} 
Olmsted {\it et al.\/} have argued that the density fluctuations resulting from
the spinodal decomposition of a polymer melt\cite{Terrill98} assist the nucleation of crystals.\cite{Olmsted98}
Strobl and coworkers have argued, on the basis of the thickness dependence of the  
crystallization and melting temperatures of syndiotactic polypropylene, and the granular 
texture in AFM images of the same polymer, that the polymer first crystallizes into
blocks, which are subsequently stabilized when they fuse into lamellae.\cite{Hugel99}
Our simulations can say little about these proposals since our polymer models are 
too simple to be able to capture such features. However, all these approaches are 
based on behaviour that has been observed in crystallization from the melt,
so it is not clear how the ideas can apply to crystallization from solution, where
the same basic laws for lamellar polymer crystals apply.

\end{multicols}

\begin{thebibliography}{10}

\bibitem{Keller57a}
A. Keller, Phil. Mag. {\bf 2},  1171  (1957).

\bibitem{Toda93}
A. Toda and A. Keller, Colloid Polym. Sci. {\bf 271},  328  (1993).

\bibitem{Keller68a}
A. Keller, Rep. Prog. Phys. {\bf 31},  623  (1968).

\bibitem{Keller96a}
A. Keller and G. Goldbeck-Wood,  in {\em Comprehensive Polymer Science, 2nd
  Supplement}, edited by S.~L. Aggarwal and S. Russo (Pergamon, Oxford, 1996),
  pp.\ 241--305.

\bibitem{Keller58}
A. Keller and A. O'Connor, Discuss. Faraday Soc. {\bf 25},  114  (1958).

\bibitem{Barham85}
P.~J. Barham {\it et~al.}, J. Mater. Sci. {\bf 20},  1625  (1985).

\bibitem{Lauritzen60}
J.~I. Lauritzen and J.~D. Hoffman, J. Res. Nat. Bur. Stds. {\bf 64},  73
  (1960).

\bibitem{Hoffman76a}
J.~D. Hoffman, G.~T. Davis, and J.~I. Lauritzen,  in {\em Treatise on Solid
  State Chemistry}, edited by N.~B. Hannay (Plenum Press, New York, 1976),
  Vol.~3, Chap.~7, p.\ 497.

\bibitem{Hoffman97}
J.~D. Hoffman and R.~L. Miller, Polymer {\bf 38},  3151  (1997).

\bibitem{Sadler84a}
D.~M. Sadler and G.~H. Gilmer, Polymer {\bf 25},  1446  (1984).

\bibitem{Sadler86a}
D.~M. Sadler and G.~H. Gilmer, Phys. Rev. Lett. {\bf 56},  2708  (1986).

\bibitem{Sadler87d}
D.~M. Sadler, Nature {\bf 326},  174  (1987).

\bibitem{Sadler88a}
D.~M. Sadler and G.~H. Gilmer, Phys. Rev. B {\bf 38},  5684  (1988).

\bibitem{Armistead92a}
K. Armistead and G. Goldbeck-Wood, Adv. Polym. Sci. {\bf 19},  219  (1992).

\bibitem{Yamamoto97}
T. Yamamoto, J. Chem. Phys. {\bf 107},  2653  (1997).

\bibitem{Yamamoto98}
T. Yamamoto, J. Chem. Phys. {\bf 109},  4638  (1998).

\bibitem{Chen98}
C.-M. Chen and P.~G. Higgs, J. Chem. Phys. {\bf 108},  4305  (1998).

\bibitem{Liu98a}
C. Liu and M. Muthukumar, J. Chem. Phys. {\bf 109},  2536  (1998).

\bibitem{Toma98}
L. Toma, S. Toma, and J.~A. Subirana, Macromolecules {\bf 31},  2328  (1998).

\bibitem{Doye98f}
J.~P.~K. Doye and D. Frenkel, J. Chem. Phys. {\bf 109},  10033  (1998).

\bibitem{Doye98d}
J.~P.~K. Doye and D. Frenkel, Phys. Rev. Lett. {\bf 81},  2160  (1998).

\bibitem{Doye99b}
J.~P.~K. Doye and D. Frenkel, J. Chem. Phys. {\bf 110},  2692  (1999).

\bibitem{Doye99d}
J.~P.~K. Doye and D. Frenkel, J. Chem. Phys. {\bf 110},  7073  (1999).

\bibitem{Doye99e}
J.~P.~K. Doye and D. Frenkel, Polymer  in press  (cond-mat/9901181).

\bibitem{Frank60}
F.~C. Frank and M. Tosi, Proc. Roy. Soc. A {\bf 263},  323  (1961).

\bibitem{Bassett62}
D.~C. Bassett and A. Keller, Phil. Mag. {\bf 7},  1553  (1962).

\bibitem{Dosiere86a}
M. Dosi\`ere, M.-C. Colet, and J.~J. Point, J. Poly. Sci. Phys. Ed. {\bf 24},
  345  (1986).

\bibitem{Siepmann92a}
J.~I. Siepmann and D. Frenkel, Mol. Phys. {\bf 75},  59  (1992).

\bibitem{Torrie77}
G.~M. Torrie and J.~P. Valleau, J. Comp. Phys. {\bf 23},  187  (1977).

\bibitem{Point79b}
J.-J. Point, Faraday Disc. Chem. Soc {\bf 68},  167  (1979).

\bibitem{Price61}
F.~P. Price, J. Chem. Phys. {\bf 35},  1884  (1961).

\bibitem{Lauritzen67b}
J.~I. Lauritzen and E. Passaglia, J. Res. Nat. Bur. Stds. {\bf 71},  261
  (1967).

\bibitem{Point79a}
J.~J. Point, Macromolecules {\bf 12},  770  (1979).

\bibitem{DiMarzio82a}
E.~A. DiMarzio and C.~M. Guttman, J. Appl. Phys. {\bf 53},  6581  (1982).

\bibitem{Point79c}
J.-J. Point, Faraday Disc. Chem. Soc {\bf 68},  366  (1979).

\bibitem{Dupirelong}
M. Dupire, PhD Thesis, {\it Etude Critique des Modeles Sequentiels de la
  Crystallization des Hauts Polymeres Lineaires}, Universit\'{e} de l'Etat
  \`{a} Mons (1984).

\bibitem{Goldbeck92a}
Golcbeck-Wood, G., PhD Thesis, {\it Computer Simulation of Polymer
  Crystallization}, University of Bristol (1992).

\bibitem{Goldbeck94a}
G. Goldbeck-Wood, Macromol. Symp. {\bf 81},  221  (1994).

\bibitem{Spinner95}
M.~A. Spinner, R.~W. Watkins, and G. Goldbeck-Wood, J. Chem. Soc., Faraday
  Trans. {\bf 91},  2587  (1995).

\bibitem{Keller93a}
A. Keller, G. Goldbeck-Wood, and M. Hikosaka, Faraday Discuss. {\bf 95},  109
  (1993).

\bibitem{Keller98}
A. Keller and S.~Z.~D. Cheng, Polymer {\bf 39},  4461  (1998).

\bibitem{Terrill98}
N. Terrill {\it et~al.}, Polymer {\bf 39},  2381  (1998).

\bibitem{Olmsted98}
P.~D. Olmsted {\it et~al.}, Phys. Rev. Lett {\bf 81},  373  (1998).

\bibitem{Hugel99}
T. Hugel, G. Strobl, and R. Thomann, Acta Polym. {\bf 50},  213  (1999).

\end{thebibliography}
\end{document}